\documentclass[twocolumn,showpacs,preprintnumbers,amsmath,amssymb,prl, reprint]{revtex4-1}
\usepackage[T1]{fontenc}
\usepackage[latin9]{inputenc}
\usepackage{amsmath}
\usepackage{graphicx}
\usepackage{hyperref}
\usepackage{colortbl}
 \usepackage[usenames,dvipsnames]{xcolor}

\begin{document}

\title{Range Expansion of Heterogeneous Populations}
\author{Matthias Reiter}\thanks{M. Reiter and S. Rulands contributed equally to this work.}
\author{Steffen Rulands}\thanks{M. Reiter and S. Rulands contributed equally to this work.}
\author{Erwin Frey}

\affiliation{Department of Physics,  Arnold-Sommerfeld-Center for Theoretical Physics and Center for NanoScience, Ludwig-Maximilians-Universit\"at M\"unchen, Theresienstrasse 37, D-80333 M\"unchen, Germany}

\begin{abstract}
Risk spreading in bacterial populations is generally regarded as a strategy to maximize survival. Here, we study its role during range expansion of a genetically diverse population where growth and motility are two alternative traits. We find that during the initial expansion phase fast growing cells do have a selective advantage. By contrast, asymptotically, generalists balancing motility and reproduction are evolutionarily most successful. These findings are rationalized by a set of coupled Fisher equations complemented by stochastic simulations.
\end{abstract}


\maketitle

Expansion of populations is a ubiquitous phenomenon in nature which includes the spreading of advantageous genes~\cite{Fisher1937} or infectious diseases~\cite{Mollison1977, Grenfell2001}, and dispersal of species into new territory. The latter has recently been investigated experimentally by analyzing the spreading of bacterial populations after droplet inoculation on an agar plate \cite{Ben-Jacob1994, Golding1999,  Beer2010, Hallatschek2007, Hallatschek2008, Korolev2010, Datta2013}. Among others, these studies have highlighted the importance of random genetic drift in driving population differentiation along the expanding fronts of bacterial colonies~\cite{Hallatschek2008, Hallatschek2007, Korolev2010, Hallatschek2010, Kuhr2011, Korolev2013}. While these studies have focused on genetically uniform populations or the competition between two strains with different growth rates~\cite{Kuhr2011,Korolev2013} much less is known about range expansion of heterogeneous populations. Single cell studies have revealed that even genetically identical bacteria exhibit variability in phenotypic traits~\cite{Dubnau2006}. As an example,  clonal populations of \emph{Bacillus subtilis} (in mid expontial growth phase) consist of both swarming cells, propelled by flagella, and non-motile cells~\cite{Kearns2005}.  Cells in the motile state do not divide. As a result, colonies of \emph{B. subtilis} are heterogeneous with respect to the cells' motility. This risk-spreading strategy allows the population to exploit nutrients at its current location and at the same time disperse to new, possibly more favorable, niches.

Motivated by these findings, we consider range expansion of a heterogeneous population. We ask what degree of risk-spreading between cell division and motility is optimal for survival during range expansion, \emph{i.e.} whether an individual is better off by investing preferentially in growth or in motility, or by adopting a risk-spreading strategy and balance its investment in growth as well as motility.
Specifically, we study range expansion dynamics on a one- and two-dimensional lattice, where each site can be occupied by an arbitrary number of individuals.  We assume that each individual $i$ has a distinct genotype $A_i$, which encodes rates to migrate, $e_i$, and reproduce, $\mu_i$, \emph{i.e.}\  in the language of game theory each individual plays a mixed strategy. In detail, an individual $A_i$ may reproduce with a rate in the interval $\mu_i\in(0,\mu_\text{max})$ upon consumption of resources $B$:  $A_i B \xrightarrow{\mu_i} A_i A_i$; the offspring inherits the genotype and is placed on the same lattice site. In addition, individuals are able to migrate upon stochastically hopping to nearest neighbor sites with a rate $e_i$ in the range $(0,e_\text{max})$. Motivated by the behavior of bacterial populations we assume that an individual may invest its limited resources partly in motility and partly in reproduction, and model this by the constraint $e_i/e_\text{max}+\mu_i/\mu_\text{max}=1$, \emph{i.e.}\ fast reproducing individuals can only move slowly and vice versa. As we will see, the implications of this biologically motivated tradeoff are more intricate than the phenomenon of front acceleration found in populations exhibiting only heterogeneous motility~\cite{Benichou2012}.  Numerical simulations of our stochastic lattice gas model were performed using Gillespie's algorithm~\cite{Gillespie1977}  with sequential updating on square and hexagonal lattices with lattice spacing $a$. We measure time in units of the inverse maximum reproduction rate $1/\mu_\text{max}$, \emph{i.e.}\ roughly speaking dimensionless time $t$ corresponds to the number of generations (of the slowest moving genotype).

\begin{figure}[htb]
\centering
\includegraphics[width=\columnwidth]{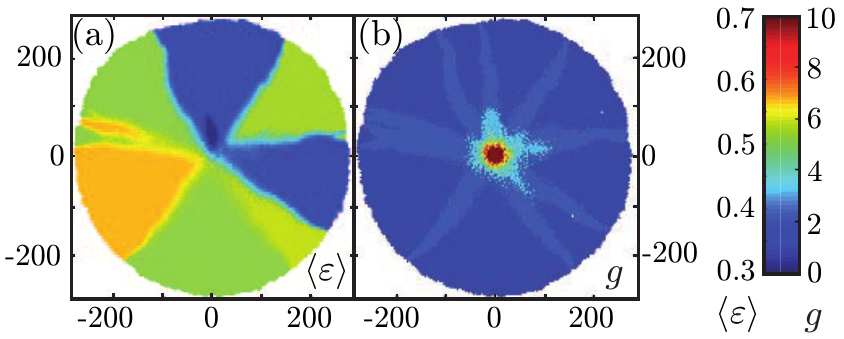}
\caption{Segregation patterns emerging from the stochastic simulation of a range expansion dynamics starting from a genetically diverse inoculum. (a) Local average motility $\langle\epsilon\rangle$ with blue (dark gray) signifying a low, yellow (light gray) a medium, and red (medium gray) a high motility. The front is dominated by genotypes with a motility close to $\epsilon =0.5$ (`generalists'). (b) Local genetic diversity with blue (dark gray) indicating a homogeneous and red  (medium gray) a heterogeneous population. Genetic diversity is rapidly lost during the range expansion process. The population remains genetically diverse only close to the inoculum and at sector boundaries. Stochastic simulations were run on a hexagonal lattice with $601\times695$ sites, with carrying capacity $\Omega=100$, initial number of genotypes $G=900$, and dimensionless time $t=490$.
\label{fig:box1}}
\end{figure}

We are interested in a range expansion scenario where initially a small area with a linear extension of three lattice sites  (inoculum) is occupied by a genetically diverse population with $G$ different genotypes $A_i$, each site containing $\Omega$ individuals, while the remainder of the lattice sites contains $\Omega$ units of resources $B$.  We assume that the local carrying capacity $\Omega$ is large: $\Omega\gg 1$ and thereby $G\gg 1$ as well. The relative hopping rates $\epsilon_i=e_i/e_\text{max}$ in the initial population are randomly drawn from a uniform distribution on $(0,1)$. Our stochastic simulations show that the inoculum quickly expands into a circular front with a concomitant loss of genetic diversity and the formation of multiple sectors composed of single genotypes; Figs.~\ref{fig:box1}(a,b) show a typical configuration in two spatial dimensions with the ensuing spatial distribution of relative motilities and genetic diversity, respectively.

\begin{figure}[htb]
\centering
\includegraphics[width=\columnwidth]{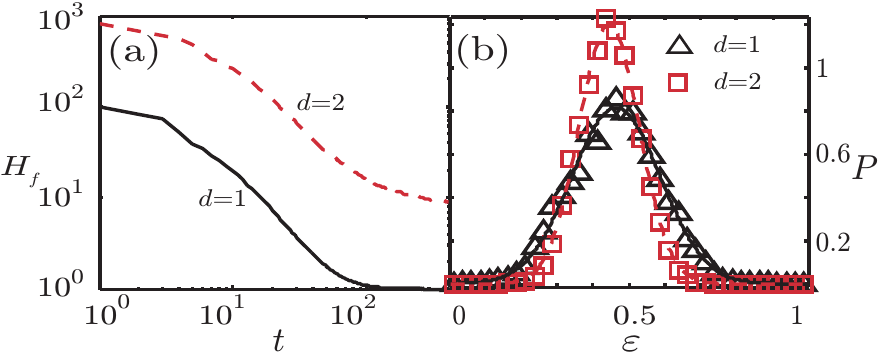}
\caption{(a) Decrease of genetic diversity $H_f(t)$ in the front region for one and two spatial dimensions. After a short transient genetic diversity decreases rapidly. While for $d=1$ genetic diversity is quickly lost, $H_f(t)=1$, it decreases only slowly in $d=2$ due to the formation of homogeneous sectors. (b) Probability to find a certain motility at a large time $t=220$. The populations is dominated by individuals with an approximately equal probability to migrate or reproduce. The histograms were averaged over $10^3$ sample runs with $\Omega=100$. 
\label{fig:box2}}
\end{figure}

The rate at which genetic diversity is lost during expansion is strongly interlinked with the underlying dynamics of the range expansion processes. Of particular importance is the genetic diversity in the front region, as these individuals constitute the gene pool for the further expansion process~\cite{Hallatschek2008}. The position of the front is defined as those lattice sites, where the fraction of resources $B$ exceeds a value of $1/2$. In polar coordinates, this yields a parametrization $r(\varphi)$ of the front, giving its distance from the origin as a function of the angle $\varphi$. Figure~\ref{fig:box2}(a) shows the time evolution of the average number $H_f(t)$ of distinct genotypes in a region $r(\varphi)\pm\Delta r$, where $\Delta r$ is proportional to the width of the front, $\Delta r\sim\ell$~\footnote{For genotypes $(\mu_\text{max},\epsilon_\text{max})$ the front width is $\ell = a \sqrt{e_\text{max}/\mu_\text{max}}$~\cite{Saarloos2003}. In our stochastic simulations we used $\delta=\sqrt{e_\text{max}/\mu_\text{max}} = 1$, and, for simplicity, choose $\Delta r = a$ [Fig.~\ref{fig:box2}(a)].}. 
We identify several temporal regimes characterized by different kinds of selection pressure acting on the individuals. After a short initial transient there is an intermediate regime, $10\lesssim t\lesssim 100$, where the loss of genetic diversity in the front region approximately follows a power law, $H_f(t)\propto t^{-\alpha}$ with $\alpha\approx 1.4 \pm 0.1$. This loss is significantly faster than for neutral evolution, where the neutral coalescence theory gives $\alpha=1$~\cite{Hinrichsen2000}. It suggests that the coalescence process is biased, meaning that some genotypes in the front region have a higher probability to go extinct than others; we will see later that this bias is related to the speed of Fisher waves for different genotypes. For $d=1$, which may for example be realized in coupled microfluidic chambers, the selection process quickly leads to the fixation of one particular genotype in the front region, as apparent from Fig.~\ref{fig:box2}(a). As opposed to this, for $d=2$, range expansion leads to the formation of monoclonal sectors with a uniform genotype [Fig.~\ref{fig:box1}(a)]. Further loss of genetic diversity is subsequently caused by annihilation of neighboring sector boundaries, and, as a result, $H_f$ decreases at a rather slow rate.

This dynamics of genetic diversity leaves two key questions: First, which genotypes are selected by the expansion process  and what is the asymptotic composition of the population?  Second, which dynamic processes lead to the asymptotic state? To answer the first question we determined the genetic composition of the population after many generations, \emph{i.e.}\  the probability $P(\epsilon)$ that an individual has a relative hopping rate $\epsilon$ and a corresponding relative reproduction rate $1-\epsilon$, [Fig.~\ref{fig:box2}(b)]. We find that successful genotypes are `generalists' which migrate and reproduce with approximately equal probability, while `specialists', who preferentially reproduce or migrate, do not colonize.
To answer the second question one needs to understand the role of evolutionary forces during range expansion. This can be achieved to a large degree within an analytical approach valid in a deterministic continuum limit where the carrying capacity, $\Omega$ is large and the local genetic diversity, \emph{i.e.}\ the number of genotypes on any lattice site, $g(\vec{r},t)$, is sufficiently low: $\Omega \gg g(\vec{r},t) \gg 1$. In the spirit of a Fisher equation~\cite{Fisher1937} one can then write down a set of coupled integro-difference-differential equations for the fraction of species with a given relative hopping rate, $n_{\epsilon_i} (\vec{r},t):=N_i (\vec{r},t)/\Omega$, and the fraction of resources, $\rho(\vec{r},t):=R (\vec{r},t)/\Omega$, where $N_i (\vec{r},t)$ and $R (\vec{r},t)$ are the local number of individuals with strategy $A_i$ and local number of resource units $B$, respectively. One obtains a set of Fisher equations for $n_{\epsilon} (\vec{r},t)$~\footnote{As each individual has a distinct genotype and $G\gg 1$ we can omit the index $i$ and formally treat $\epsilon$ as a continuous variable uniquely identifying a certain genotype.} coupled through the availability of resources $\rho (\vec{r},t)$:
\begin{subequations} \label{eq:ipde}
\begin{align} 
\partial_t n_\epsilon(\vec{r},t) &=D_\epsilon \Delta n_\epsilon(\vec{r},t) + (1-\epsilon) n_\epsilon(\vec{r},t) \rho(\vec{r},t) \, , \\
\partial_t\rho(\vec{r},t)&=-\rho (\vec{r},t)\int_0^1\! (1-\epsilon)\, n_\epsilon(\vec{r},t)\,\mathrm{d}\epsilon\, .
\end{align}
\end{subequations}
Here $\Delta$ is the lattice Laplacian, $D_\epsilon=\epsilon/(2d\delta^2)$ with $\delta=\sqrt{\mu_\text{max}/e_\text{max}}$, and the unit of length is $\ell=a/\delta$. 

Equations~\ref{eq:ipde}(a,b) exhibit a stationary, spatially uniform state with resources only:  $n(\vec{r},t)\equiv\int_0^1\! n_\epsilon(\vec{r},t)\,\mathrm{d}\epsilon=0$ and $\rho(\vec{r},t)=1$. However, a linear stability analysis shows that this state is locally unstable to small population seeds. The ensuing exponential growth is limited by the availability of resources, and saturates when resources are fully exploited, $\rho(\vec{r},t)=0$, and the population reaches the carrying capacity, $n(\vec{r},t)=1$. From classical front propagation theory we expect that a small population seed will develop into a traveling wave front~\cite{Saarloos2003}. Indeed, in accordance with our stochastic simulations, a numerical solution of Eqs.~\ref{eq:ipde}(a,b) shows  propagating wave solutions. For a homogeneous system with growth rate $\mu_{\text{max}}$ and migration rate $e_\text{max}$ the width of such front is $\ell=a/\delta$. Hence $\delta$ measures the ``coarseness'' of the model: It can be read as the size of a bacterium, $a$, compared to the width of the wave front, $\ell$. Alternatively, $\delta=\sqrt{\mu_\text{max}/e_\text{max}}$ also gives the relative maximal range of growth and hopping rates. While for small values of $\delta$ diffusion is faster than growth, large values correspond to a growth-dominated expansion process.

\begin{figure}[b] 
\centering
\includegraphics[width=\columnwidth]{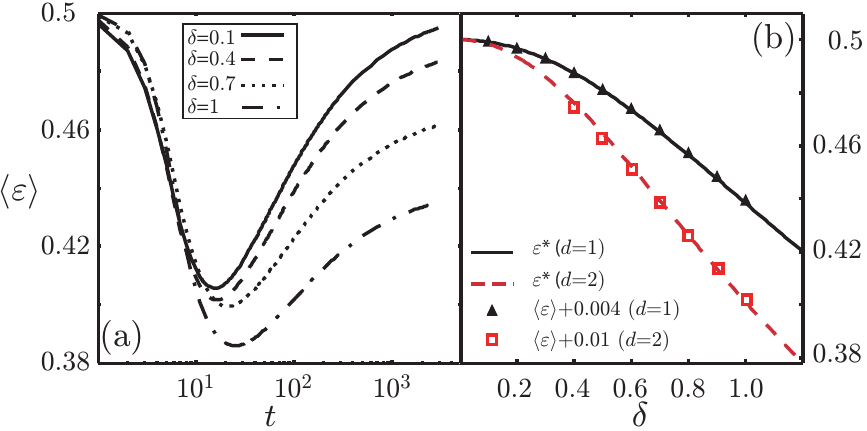}
\caption{\label{fig:box3}(a) To investigate which genotypes are selected by the evolutionary dynamics at specific times we numerically solved the Fisher equations, Eqs.~\ref{eq:ipde}(a,b), for various values of $\delta$ (as indicated in the graph) and computed the average motility $\langle\epsilon\rangle$ in the population. (b) The solid (dashed) line illustrates the analytical result for the genotype $\epsilon^*$  with the optimal front velocity in one (two) spatial dimensions, [Eq.~\eqref{eq:eps}]. Triangles indicate the average genotype $\langle\epsilon\rangle$ obtained from the numerical solution of Eqs.~\eqref{eq:ipde}, evaluated at a large time $t=3000$ for $d=1$ and at $t=1100$ for $d=2$, respectively.}
\end{figure}
To understand the role of evolutionary forces during range expansion we computed the temporal evolution of the mean motility $\langle\epsilon\rangle$ in the whole population. Figure~\ref{fig:box3}(a) shows $\langle\epsilon\rangle (t)$  as obtained from the numerical solution of the coupled Fisher equations, Eqs.~\ref{eq:ipde}(a,b), for a series of values for $\delta$. 
During the first few generations, $t\lesssim 15$, while the population is genetically still highly heterogeneous, the population dynamics is governed by scramble competition for resources.  In order to dominate the front, a potentially successful genotype must be capable of efficiently outgrowing its competitors by consumption of the majority of resources at the front. This gives a selective advantage to genotypes with a high reproduction rate. They dominate over competitors with a higher motility, which in turn leads to a decrease in the mean motility $\langle\epsilon\rangle$, cf. Fig.~\ref{fig:box3}(a); the decrease is to a large degree independent of $\delta$. After this initial phase,  for $t\gtrsim15$, macroscopic differences in the concentrations of the different genotypes have emerged which locally compete for resources. Our simulations show that the average motility reaches a minimum and starts to increase again [Fig.~\ref{fig:box3}(a)].  This indicates that now the evolutionary most successful genotypes are no longer those which optimize their growth rates (specialists), but those, which balance reproduction with motility. The reason is that the decisive factor limiting the growth of a particular genotype colony is the velocity of the ensuing Fisher wave, as can be understood by analyzing the set of coupled Fisher equations, Eqs.~\ref{eq:ipde}(a,b): Since the velocity of a Fisher wave is determined by its leading edge where resources are plentiful, we may approximately write $\rho\approx 1-n$. Following the theory of front propagation~\cite{Brunet1997, Saarloos2003}, we assume that traveling wave solutions $n_\epsilon(r,t)=n_\epsilon(r-vt)=n_\epsilon(z)$  decay exponentially at the leading edge of the front, $n_\epsilon(z)\sim \exp{(-\gamma z)}$. Upon substituting the exponential ansatz into Eqs.~\ref{eq:ipde}(a,b) and linearizing in the concentrations we find that the Fisher equations for different values of $\epsilon$ decouple~\cite{Supplement}. Keeping only the highest order exponential terms, we obtain a dispersion relation $v(\gamma)=\gamma^{-1}\left\{(\epsilon/d) \left[\cosh(\delta\gamma)-1\right]+1-\epsilon\right\}$. Given a sufficiently steep initial front, the solution with minimal velocity $v(\gamma_0)$ is selected~\cite{Bramson1983, Saarloos2003};  for a radially expanding front in $d=2$, $v(\gamma_0)$ is approached asymptotically for $r\to\infty$~\citep{Murray2002}. Hence, a homogeneous subpopulation with motility 
\begin{equation}
\epsilon^*=\left[\sqrt{2/(d\delta^2)+1}\;\mathrm{arccosh}\left(1+d\delta^2\right)\right]^{-1}\label{eq:eps}
\end{equation}
has the highest invasion speed. As illustrated in Fig.~\ref{fig:box3}(b), the optimal motility is $\epsilon^*=0.5$ for $\delta\to 0$, and it decreases only slowly with increasing $\delta$. Since the fastest propagating subpopulation will take an increasingly larger fraction of the colony, this explains why the mean motility increases [Fig.~\ref{fig:box3}(a)]. Concomitant with this coarsening process sectors of uniform genotypes form for $50\leq t\leq 100$; see Fig.~\ref{fig:box1}(a,b).

For a radially expanding front in two dimensions, the subsequent population dynamics is mainly governed by annihilation of these sector boundaries. Since this is a very slow process, the mean motility $\langle\epsilon\rangle$ is only asymptotically approaching the optimal value $\epsilon^*$: The boundary $\varphi(r)$ of two adjacent sectors, propagating with velocities $v_1$ and $v_2>v_1$, forms a logarithmic spiral with $\varphi(r)=-\sqrt{v_2^2/v_1^2-1}\,\ln(r/a)$ which moves only very slowly into the direction of the slower domain~\footnote{Neglecting fluctuations the angle $\varphi$  satisfies the differential equation~\citep{Korolev2012}, $r\varphi'(r) =-\sqrt{v_2^2/v_1^2 - 1}$, which is solved by  $\varphi(r)=-\sqrt{v_2^2/v_1^2-1}\ln(r/a)$ where we have taken without loss of generality $\varphi(a) =0$.}. Hence any front consisting of multiple sectors will ultimately be dominated by the genotype with the fastest front velocity, given by Eq.~\eqref{eq:eps}. However, this annihilation process is far too slow to be observed within the numerically accessible time window. To heuristically account for this, we have added a constant value to $\langle\epsilon\rangle$. We find  excellent agreement of $\epsilon^*$ and $\langle\epsilon\rangle$ strongly suggesting that the asymptotic genotype is determined by the optimal velocity of the corresponding homogeneous fronts.

To investigate the dependence of the asymptotic composition of the population on the strength of reaction noise, we computed $\langle\epsilon\rangle$ and $\epsilon^*$ also by employing stochastic simulations of the lattice gas model for different values of the system size $\Omega$.  Demographic noise affects the evolutionary dynamics in manifold ways: The initial coarsening process leading to sector formation is an inherent stochastic process which is not well accounted for by the set of Fisher equations, Eqs.~\ref{eq:ipde}(a,b). Figure 4 illustrates that compared to the deterministic dynamics the coarsening process for the stochastic dynamics is slightly faster.
\begin{figure}[tb] 
\centering
\includegraphics[width=\columnwidth]{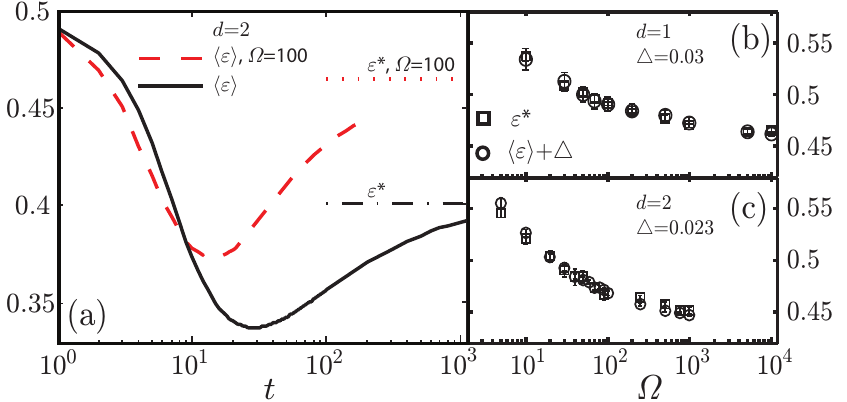}
\caption{\label{fig:box4}(a) Comparison of $\langle\epsilon\rangle(t)$ between stochastic simulations with $\Omega=100$ and simulations of Eqs.~\eqref{eq:ipde}, both for $\delta=1$ and $d=2$. The measured value $\epsilon^*$ for $\Omega=100$ from (c), and the analytic value from  Eq.~\eqref{eq:eps} are indicated. (b, c) Comparison of $\langle\epsilon\rangle$ as obtained from stochastic simulations at a large time $t=220$ with $\epsilon^*$ as function of $\Omega$. These quantities differ by a small constant value $\Delta$ indicated in the graph. Stochastic simulation results in (a-c) were averaged over at least $500$ sample runs.}
\end{figure}
Moreover, the ensuing sector boundaries also merge faster due to the stochastic meandering motion of the sector boundaries~\cite{Ali2010,Hallatschek2007}. Indeed we recover that the stochastic lateral movement of domain boundaries is super-diffusive, \emph{i.e.}\  its root-mean-square displacement increases as $t^\gamma$, with $\gamma>0.5$.  For a sector boundary of a planar front we measured $\gamma\approx 0.63$ \cite{Supplement} confirming that conformation of the sector boundaries is well described by kinetic roughening~\cite{Kardar1986, Saito1995}. These stochastic effects become less important as the colony grows~\cite{Ali2010,Korolev2010,Lavrentovich2013,Ali2013}: Since the front is advancing uniformly, \emph{i.e.}\  $t\sim r$, the front roughness $r^\gamma$ becomes small compared to deterministic drift $r\ln r$ as $r\to\infty$. Conversely, due to the absence of front inflation, sector annihilation proceeds more rapidly as a result of stochastic fluctuations for planar fronts~\cite{Supplement,Korolev2010,Ali2010,Ali2013,Lavrentovich2013}. 
	
Finally, noise also affects the speed of propagating fronts~\cite{Brunet1997, Panja2004, Brunet2006, Hallatschek2009, Hallatschek2011, Hallatschek2011a}. Taken together, we find that demographic noise significantly affects the population dynamics during range expansion and leads to an asymptotic composition of the population with an  increased average motility, cf. Fig.~\ref{fig:box4}(a,b). In particular, the asymptotic value of $\langle\epsilon\rangle$ and the genotype $\epsilon^*$ with the highest velocity of homogeneous fronts both decrease with $\Omega$. In fact, $\langle\epsilon\rangle$ and $\epsilon^*$ differ only by a small constant, which can be attributed to the fact that $\langle\epsilon\rangle$ was measured at a finite time. This observation underscores our assertion that the species dominating the front will be the genotype maximizing its front speed.

In conclusion, we studied the role of risk-spreading between motility and growth during range expansion. Starting from a genetically heterogeneous population we found that during the initial phase of the expansion process scramble competition for resources favors fast growing individuals. Concomitantly, the number of distinct genotypes decreases rapidly and thereby genetically homogeneous sectors form. Therefore, the competitive advantage at larger times shifts towards those individuals with the highest front speed. We have shown that risk-spreading leads to an optimal front speed. In the deterministic limit, described by a set of coupled Fisher equations, the optimal strategy turns out to be perfect risk-spreading between  
motility and growth in a parameter regime dominated by diffusion (small dimensionless parameter $\delta$). Our analytical results also quantify how the optimal strategy is increasingly biased towards growth as the typical time scales for growth and diffusion become comparable. A low carrying capacity is affecting the range expansion dynamics in a twofold way: During the initial phase demographic noise may lead to an early fixation of the front and hence to a bias towards slowly migrating individuals. At later stages of range expansion, noise leads to a strong shift of the optimal value for the mean motility towards larger values. 

We expect that both the spatial separation of different genotypes and the evolutionary success of generalists are generic for range expansions of heterogeneous populations. By genetically tuning the number of flagella in \emph{E. coli} bacteria, a motility-growth tradeoff can be studied experimentally. Current experiments are investigating the implications of this tradeoff for range expansion, allowing for a test of our results~\cite{Taute2014}. We believe that our model can also be tested using custom-build reaction-diffusion networks with synthetic nucleic acids and enzymatic reactions \cite{Padirac2013}. In general, our findings also pertain to other spreading processes, where motility is complementary to growth. Further work might include mutations~\cite{Kuhr2011,Greulich2012} or extend our findings to excitable media, systems exhibiting an Allee effect~\cite{Taylor2005} and metastable states~\cite{Meerson2011,Rulands2013}, and finally also to more complex reaction networks~\cite{Reichenbach2008, Case2010, Dobrinevsky2012, Knebel2013}. \\
 
\begin{acknowledgments}
We thank Madeleine Opitz and David Jahn for fruitful discussions. This research was supported by the Deutsche Forschungsgemeinschaft via contract no. FR 850/9-1 and the German Excellence Initiative via the program  `Nanosystems Initiative Munich'. S.R. gratefully acknowledges support of the Wellcome Trust (grant number 098357/Z/12/Z).
\end{acknowledgments}

%

\end{document}